\documentclass[a4paper,11pt]{article}
\pdfoutput=0 

\usepackage{jheppub} 

\usepackage[T1]{fontenc} 

\title{\boldmath Statistical issues in the parton
distribution analysis of the Tevatron jet data} 
\preprint {\scriptsize DESY-12-191,~DO-TH~12/34,~LPN12-118,~SFB/CPP-12-79
\normalsize}

\author[a,b,1]{S.Alekhin,\note{Corresponding author.}}
\author[a]{J.Bl\"umlein,}
\author[a,c]{S.-O.Moch}

\affiliation[a]{DESY, Platanenallee 6, D--15738 Zeuthen, Germany}
\affiliation[b]{Institute for High Energy Physics,                            
    142281 Protvino, Moscow region, Russia}
\affiliation[c] {II. Institut f\"ur Theoretische Physik, Universit\"at Hamburg, Luruper Chaussee 149, D-22761 Hamburg, Germany}

\emailAdd{sergey.alekhin@desy.de}
\emailAdd{johannes.bluemlein@desy.de}
\emailAdd{sven-olaf.moch@desy.de}

\abstract{We analyse a tension between the D0 and CDF  
inclusive jet data and the perturbative QCD calculations, which are based on the
ABKM09 and ABM11 parton distribution functions (PDFs)
 within the nuisance parameter framework.
Particular attention is paid on the uncertainties in the 
nuisance parameters due to the data fluctuations and the PDF errors. 
We show that with account of these uncertainties the nuisance parameters 
do not demonstrate a statistically significant excess. A statistical bias 
of the estimator based on the nuisance parameters is also discussed. 
}

\begin{document} 
\maketitle
\flushbottom

\section{Introduction}
\label{sec:intro}

Since the first observation of jet production at Tevatron this process 
is considered as a valuable source of information about the gluon distribution 
at large $x$. Indeed, the gluon distribution directly enters into the 
jet production cross section in contrast to the deep-inelastic-scattering 
(DIS) process, which  
provides only an indirect constraint on the gluon distribution, through the QCD 
evolution. The Tevatron jet production
data~\cite{Abazov:2008ae,Abulencia:2007ez} are used in the global
fits of parton distribution functions (PDFs) 
to improve accuracy of the gluon distribution, particularly at large $x$.  
At this end proper statistical treatment of the data is required
since uncertainties in the data of 
Refs.~\cite{Abazov:2008ae,Abulencia:2007ez} are 
dominated by the correlated systematics and the simplest $\chi^2$ 
estimator is inapplicable. In this case one should ideally use 
the $\chi^2$ estimator including the covariance matrix, which
encodes the error correlations. However, for the sake of 
implementation simplicity an alternative form of estimator is often 
employed~\cite{Stump:2001gu}. This form is based on the so-called ``nuisance''
parameters, which describe a possible shift of the data due to systematic 
uncertainties. The nuisance parameters entering the estimator  
are fitted to the data simultaneously with 
other parameters describing the PDF shape. As a result, the number of fitted 
parameters dramatically grows. This difficulty is circumvented 
because the nuisance parameters enter into the estimator
of Ref.~\cite{Stump:2001gu} 
linearly therefore the $\chi^2$ value can be minimized with respect to 
the nuisance parameters analytically. As an added feature, 
the approach based on the nuisance parameters allows for the  
visualization of any tension between the data and the fitted model since 
it shows how large a shift of the data provides the best agreement with the
model. Moreover, in the same way the best values of the nuisance 
parameters can be estimated for any given data set, 
which is not included in the PDF fit, in order to check for 
potential problems with accommodation of the new data into the fit.

The ABKM09 PDFs~\cite{Alekhin:2009ni} and their refined version, 
ABM11 PDFs~\cite{Alekhin:2012ig}, were extracted to 
next-to-next-to-leading-order
(NNLO) in perturbative QCD from a 
combination of the world inclusive DIS data supplemented by the fixed-target
data for the Drell-Yan process and dimuon production in the neutrino-nucleon 
collision.
The Tevatron jet data were also included into a variant of the ABKM09 
fit~\cite{Alekhin:2009ni,Alekhin:2011cf} 
and good agreement with other data used in the fit has 
been achieved. The analysis of Ref.~\cite{Alekhin:2011cf} is focused on the
impact of the Tevatron data on the Higgs cross section estimate,
cf. also~\cite{Alekhin:2010dd}, and 
statistical aspects in this analysis have not been detailed. 
In the present paper we fill 
this gap by giving a detailed calculation of the nuisance parameters 
for the ABKM09 and ABM11 PDFs with and without Tevatron jet data included. 
The paper is organized as follows. In 
Section~\ref{sec:basics} we give a brief outline of the formalism used 
in analysis of the correlated data. Section~\ref{sec:results} contains a 
description of the systematic uncertainties in the Tevatron jet data and the 
corresponding nuisance parameters in comparison with ones obtained 
with other PDF sets. Particular attention is payed on the nuisance 
parameters for the normalization uncertainty and 
on the impact of this source of uncertainty on the fit results
following suggestions of Ref.~\cite{Thorne:2011kq}.  
Section~\ref{sec:concl} contains a conclusion.  

\section{Basics of the correlated data analysis}
\label{sec:basics}

In case measurements are subject to correlated systematic 
uncertainties the experimental data $\{y_i\}$ can be represented as 
follows,
\begin{equation}
y_i=f_i(\vec{\Theta}) + \mu_i \sigma_i 
+\sum_{k=1}^{N_{syst}}\lambda_k s_{k,i}^{rel} f_i(\vec{\Theta}),
\label{eq:data}
\end{equation}
where $f_i$ is the mathematical expectation of the measurement $i$
depending on the vector of model parameters $\vec{\Theta}$, $\sigma_i$ is
its uncorrelated uncertainty, $s_{k,i}^{rel}$ are the relative 
correlated uncertainties, which 
stem from $N_{syst}$ independent sources, and the index $i$ runs 
over all experimental data points. The independent random variables 
$\mu_i$ and $\lambda_k$ describe the uncorrelated and correlated fluctuations
in the data, respectively. By definition, the uncorrelated fluctuations are 
independent for each data point. In contrast, the correlated fluctuation due 
to each source $k$ are common for all data points. Routinely they are related
to systematic effects in the data normalization, calibration, corrections,
etc. For cross section measurements these factors are applied to the
data multiplicatively therefore the systematic errors are commonly 
multiplicative. With account of the 
data correlations the $\chi^2$-estimator reads
\begin{equation}
\chi^2=\sum_{ij}(y_i-f_i) E_{ij}  (y_j - f_j).
\label{eq:chi2}
\end{equation}
The error matrix $E_{ij}$ is the inverse of the positive definite 
covariance matrix $C_{ij}$. 
For the model of Eq.~(\ref{eq:data}) it reads  
\begin{equation}
C_{ij}=\sigma_i^2 \delta_{ij} 
+ \sum_{k=1}^{N_{syst}} s_{k,i}^{rel} s_{k,j}^{rel} f_i f_j,
\label{eq:covm}
\end{equation}
 where $\delta_{ij}$ is the Kronecker symbol.
Alternatively, the error correlations 
are often taken into account employing the following form of 
$\chi^2$~\cite{Stump:2001gu} 
\begin{equation}
\chi^2=\sum_i \frac{[f_i-(1 - \sum_k \eta_k s_{k,i}^{rel}) y_i]^2}{\sigma_i^2} 
+ \sum_{k=1}^{N_{syst}}\eta_k^2.
\label{eq:chi2w}
\end{equation}
The form of Eq.~(\ref{eq:chi2w}) allows for shifts of the data by the correlated
uncertainty scaled with the values of the parameters $\eta_k$. The latter  
are fitted simultaneously with the theoretical model 
parameters $\vec{\Theta}$ and in this way describe the data shifts, which 
provide the best description of the fitted model.
The form of Eq.~(\ref{eq:chi2w}) corresponds to the case, when the correlated  
uncertainties are additive, i.e. the statistical model of the data looks like
\begin{equation}
y_i^{add}=f_i(\vec{\Theta}) + \mu_i \sigma_i 
+\sum_{k=1}^{N_{syst}}\lambda_k s_{k,i}
\label{eq:dataadd},
\end{equation}
where $s_{k,i}=s_{k,i}^{rel} y_i$ and
the covariance matrix, which should be used in Eq.~(\ref{eq:chi2}) 
for this data model, reads 
\begin{equation}
\left(C^{add}\right)_{ij}=\sigma_i^2 \delta_{ij} + \sum_{k=1}^{N_{syst}} s_{k,i}
s_{k,j}.
\label{eq:cova}
\end{equation}
The advantage of the estimator in Eq.~(\ref{eq:chi2w}) is essentially
its technical simplicity
since the vector of $\eta_k$, which provides the 
minimum of Eq.~(\ref{eq:chi2w}) 
can be found analytically as a product of two matrices 
\begin{equation}
r_k=\sum_{k^\prime=1}^{N_{syst}} A_{~~k k^\prime}^{-1} B_{k^\prime},
\label{eq:nuis}
\end{equation}
where 
\begin{equation}
 A_{k k^\prime}=\delta_{k k^\prime} +
\sum_i \frac{s_{k,i} s_{k^\prime,i}}{\sigma_i^2} 
\label{eq:nuia}
\end{equation}
and 
\begin{equation}
 B_{k^\prime}=\sum_i \frac{(y_i-f_i)}{\sigma_i^2}s_{k^\prime,i}. 
\label{eq:nuib}
\end{equation}
The value of the estimator in Eq.~(\ref{eq:chi2w}) at $\eta_k=r_k$ reads
\begin{equation}
\chi^2_{min}=\sum_i \frac{(f_i-y_i)^2}{\sigma_i^2}
 - \sum_{k=1}^{N_{syst}} r_k B_k.
\end{equation}
Since the inverse of the additive covariance matrix Eq.~(\ref{eq:cova}) is 
\begin{equation}
\left(C^{add}\right)^{-1}_{i,j}=\frac{\delta_{ij}}{\sigma_i^2}  - \frac{1}{\sigma_i^2 \sigma_j^2}\sum_{k,k^{\prime}=1}^{N_{syst}} {s_{k,i} s_{k^{\prime},j}}A^{-1}_{~~k k^{\prime}}
\end{equation}
the value of $\chi^2_{min}$ 
coincides with the one of Eq.~(\ref{eq:chi2}) for the 
statistical model of data with additive systematic errors.  
The nuisance parameters $r_k$ are random variables with average equal 
to zero and the variances, which read 
\begin{equation}
V(r_k)=\sqrt{\sum_{l l^\prime}A_{~~k l}^{-1}C^{B}_{l l ^\prime}A_{~~l ^\prime k}^{-1}} 
\label{eq:nuisd}
\end{equation}
where
\begin{equation}
C^{B}_{l l^\prime}=\sum_{ij} s_{l,i} s_{l ^\prime ,j} \frac{C^{add}_{ij}}
{\sigma_i^2 \sigma_j^2}
\end{equation}
is the covariance matrix for the vectors $B_{l,l^\prime}$ of
 Eq.~(\ref{eq:nuib})\footnote{Note that the
variances of nuisance parameters differ from the square root of 
the diagonal elements 
of the inverse Hessian for Eq.~(\ref{eq:chi2w})  
equal to  ${A_{~~k k}^{-1}}$.}.
Through $f_i(\Theta)$ entering Eq.~(\ref{eq:nuib}) the nuisance parameters  
depend on the fitted parameters $\vec{\Theta}$. For the data sets,
which are not included into the fit, the nuisance 
parameters are generally bigger than the ones 
obtained from a fit, which includes those data sets,
due to better a tuning of $\vec{\Theta}$ to the data in the latter case. 
In the following Section we analyze this trend for the 
different Tevatron jet data with respect to 
the ABKM09~\cite{Alekhin:2009ni} and ABM11~\cite{Alekhin:2012ig} fits
considering two cases: 
before and after these data are included into the fit.  

\section{The Tevatron jet data in the ABKM09 and ABM11 fits}
\label{sec:results}

The Tevatron experiments CDF and D0 have accumulated big samples of 
events with hard jets in the final state and have performed elaborated 
analyses of these samples with different jet definition 
algorithms, cf.~\cite{Wobisch:2012iu} 
for a recent review. For brevity we consider 
in the following only two Tevatron inclusive jet data 
sets~\cite{Abazov:2008ae,Abulencia:2007ez} obtained
by the D0 and CDF collaborations, respectively, which nonetheless 
give a representative illustration of the issues discussed in the paper. 
\begin{figure}[tbp]
\centering 
\includegraphics[width=.9\textwidth]{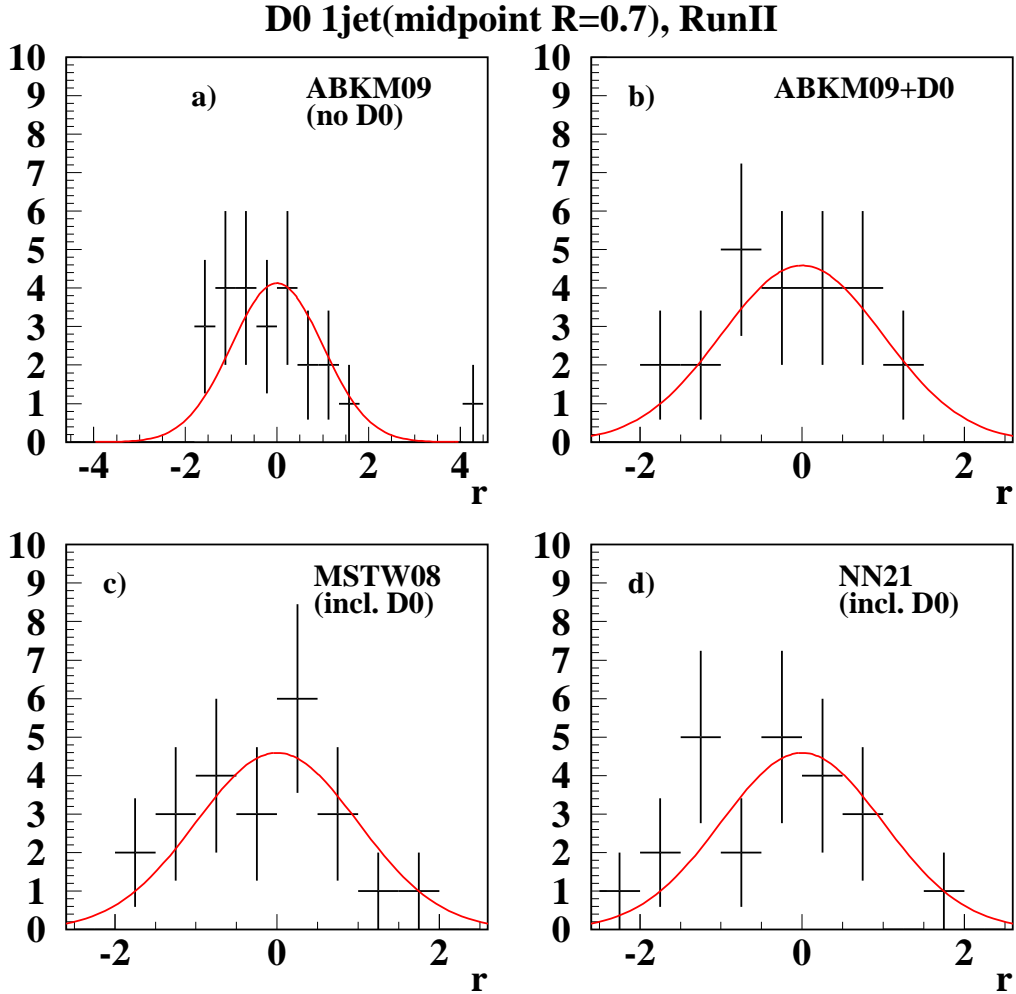}
\hfill
\caption{\label{fig:nuisd0} The distribution of nuisance parameters 
$r$ for the 
D0 data~\cite{Abazov:2008ae} on the inclusive jet production 
calculated with the threshold NNLO corrections taken into account 
and different NNLO PDFs (a): ABKM09~\cite{Alekhin:2009ni}; 
b): variant of ABKM09 obtained from 
the fit with the D0 data included~\cite{Alekhin:2011cf}; 
c): MSTW08~\cite{Thorne:2008am}; 
d): NN21~\cite{Ball:2011uy}). The curves superimposed display a 
normal Gaussian distribution normalized on the total number of the nuisance 
parameters.}
\end{figure}
Both data sets were collected in Run~II and each corresponds to an
 integral luminosity of about 1fb$^{-1}$.

The D0 analysis of Ref.~\cite{Abazov:2008ae} is based on the midpoint 
cone algorithm for the jet definition. 
\begin{figure}[tbp]
\centering 
\includegraphics[width=.9\textwidth,height=.8\textwidth]{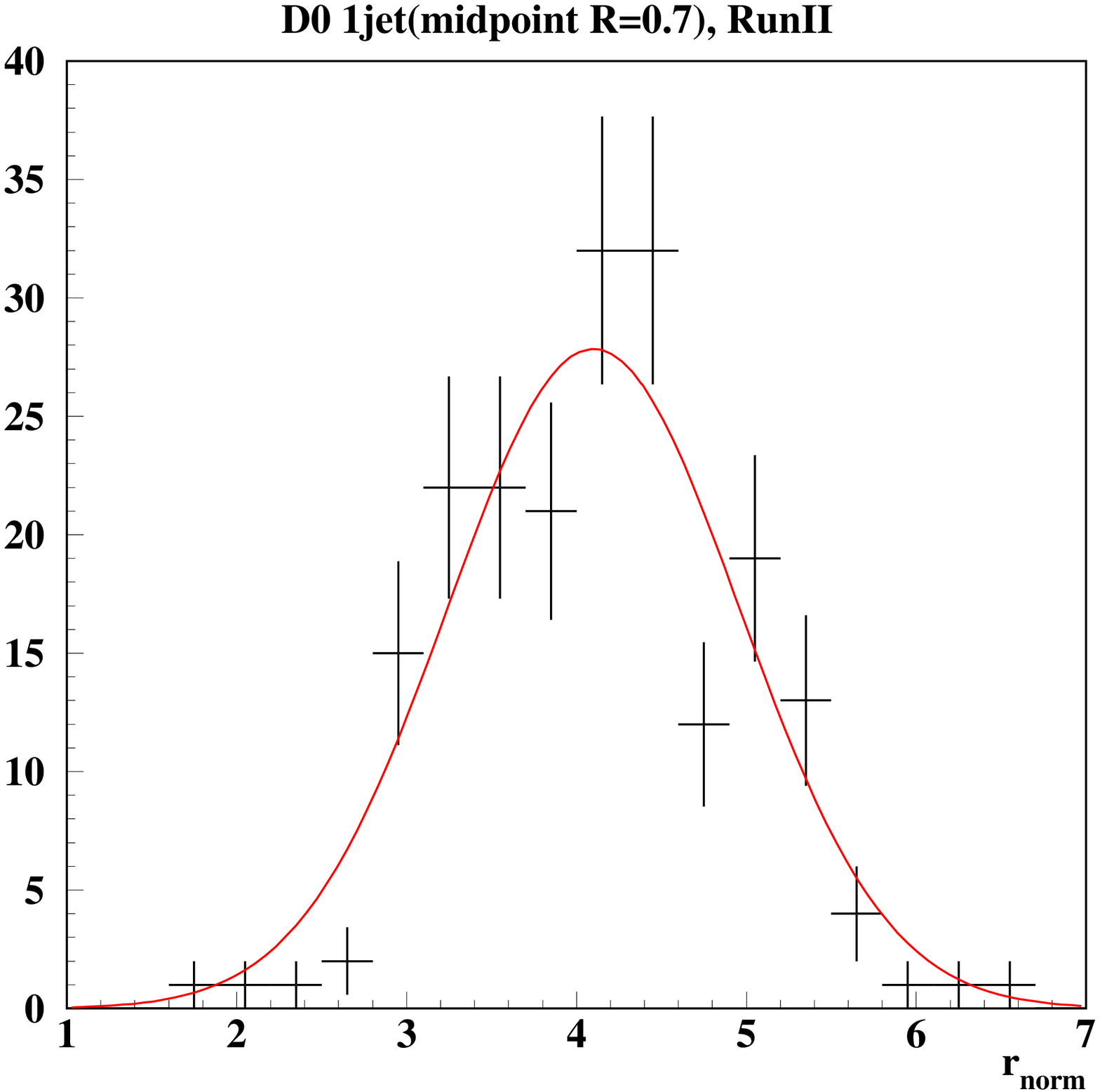}
\hfill
\caption{\label{fig:dispd0} Distribution of the D0 normalization nuisance 
parameter obtained for 200 pseudo-data sets. The curve superimposed displays
a Gaussian distribution with the average of Eq.~(\ref{eq:nuis}) and 
the variance of Eq.~(\ref{eq:nuisd}).
}
\end{figure}
\begin{figure}[tbp]
\centering 
\includegraphics[width=.49\textwidth]{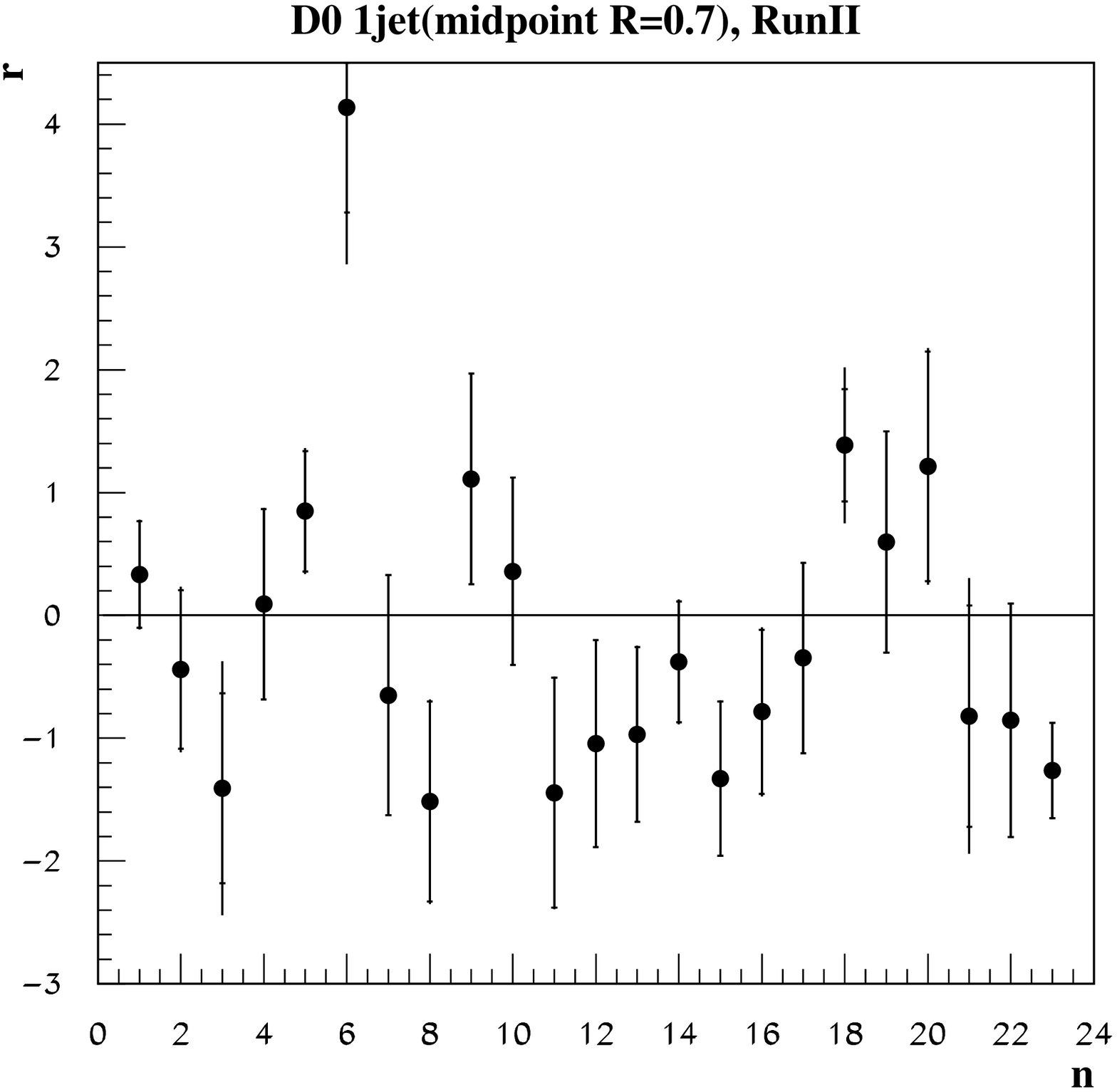}
\includegraphics[width=.49\textwidth]{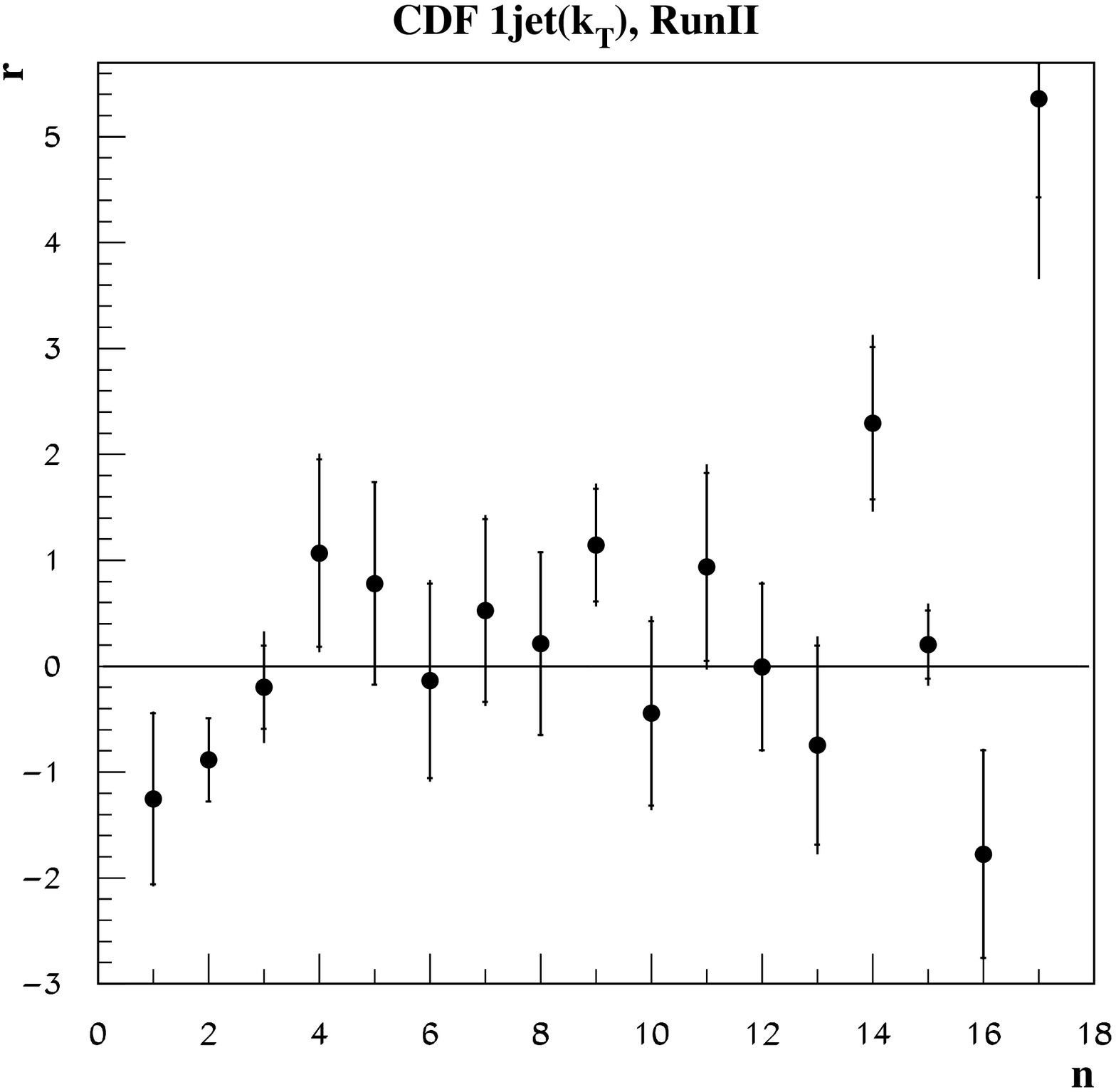}
\hfill
\caption{\label{fig:errnd0} Values of the nuisance parameters $r$
for the D0 data~\cite{Abazov:2008ae} (left) and 
the CDF ones~\cite{Abulencia:2007ez} (right)
with the uncertainties due to data fluctuation (inner bars) and 
the total uncertainties including the ones due to PDFs (outer bars) versus
the nuisance parameter number $n$. The normalization nuisance parameters 
correspond to $n=6$ and 17 for D0 and CDF, respectively. 
}
\end{figure}
The D0 data 
cover the range of $-2.4\div 2.4$ in the jet rapidity and $50\div 600$ GeV
in the transverse momentum of jet. The published correlated  
systematic uncertainties in the D0 data are due to the global normalization 
and 23 additional sources, including the jet energy calibration, resolution,
etc. In the present analysis we consider all these sources taking the average 
in the case of asymmetric errors.\footnote{The experimental data tables  
used in the analysis are available from {\tt http://arxiv.org}
as an attachment to the arXiv version of our paper.}
The distribution of the nuisance parameters $r$ of Eq.~(\ref{eq:nuis}),
which correspond to these 23 sources of systematics, calculated for 
the NNLO ABKM09 PDFs are given in Fig.~\ref{fig:nuisd0}. The jet production 
cross sections are obtained with the FastNLO tool~\cite{Kluge:2006xs}
and include the NLO corrections~\cite{Nagy:2001fj,Nagy:2003tz} 
and the threshold resummation corrections
of Ref.~\cite{Kidonakis:2000gi}. 
The D0 nuisance parameters spread in the range from 
-1.5 to 4.1 and in general their distribution is 
comparable to the normal Gaussian one.  
The maximal absolute value of $r$ corresponds to 
the systematic uncertainty in the general normalization. This
reflects the fact that the D0 data systematically overshoot the 
ABKM09 predictions, cf. Refs.~\cite{Alekhin:2012ig,Alekhin:2011cf}.
However with account of the errors in the 
nuisance parameters due to fluctuations in the data and due to the PDF
uncertainties the statistical significance of the spread in the 
nuisance parameters reduces.
To check in details the uncertainty in the D0 normalization nuisance
parameter due to the data fluctuation we calculate it for 
200 pseudo-data sets generated with Eq.~(\ref{eq:dataadd}) and the data 
errors of Ref.~\cite{Abazov:2008ae} taking a normal 
Gaussian distribution for the random variables 
$\mu$ and $\lambda$. The distribution of the normalization nuisance parameter  
obtained for these data sets 
is displayed in Fig.~\ref{fig:dispd0}. It is comparable to
the Gaussian distribution with the average of 
Eq.~(\ref{eq:nuis}) and variance of Eq.~(\ref{eq:nuisd}), which 
are $r_{norm}= 4.1$ and $V(r_{norm})=0.85$, respectively. 
The error in the nuisance parameters due to PDFs is estimated in our 
analysis as a combination of their variation with the 
change in the PDFs between the central value and each of the 25 
PDF sets describing the ABKM09 PDF uncertainties.
For the D0 nuisance normalization parameter this gives an 
additional uncertainty of $\Delta^{PDF}( r_{norm})= 0.95$. 
A combination of $V(r_{norm})$
and $\Delta^{PDF}( r_{norm})$ in quadrature gives the total uncertainty  
$\Delta^{tot}( r_{norm})=1.3$.
With account of these uncertainties the D0 normalization nuisance 
parameter is consistent with zero within 3 standard deviations.
Other D0 nuisance parameters are also consistent with 
zero within uncertainties, cf. Fig.~\ref{fig:errnd0}, therefore the statistical 
significance of the excess in the normalization nuisance parameter 
is marginal. Indeed, in the variant 
of the ABKM09 fit with the D0 data included the nuisance parameters are 
in general much smaller due to better tuning of the PDFs to the data and 
the value of normalization nuisance 
parameter is 1.5 only that is consistent with zero within the errors. 
\begin{figure}[tbp]
\centering 
\includegraphics[width=.98\textwidth,height=.6\textwidth]{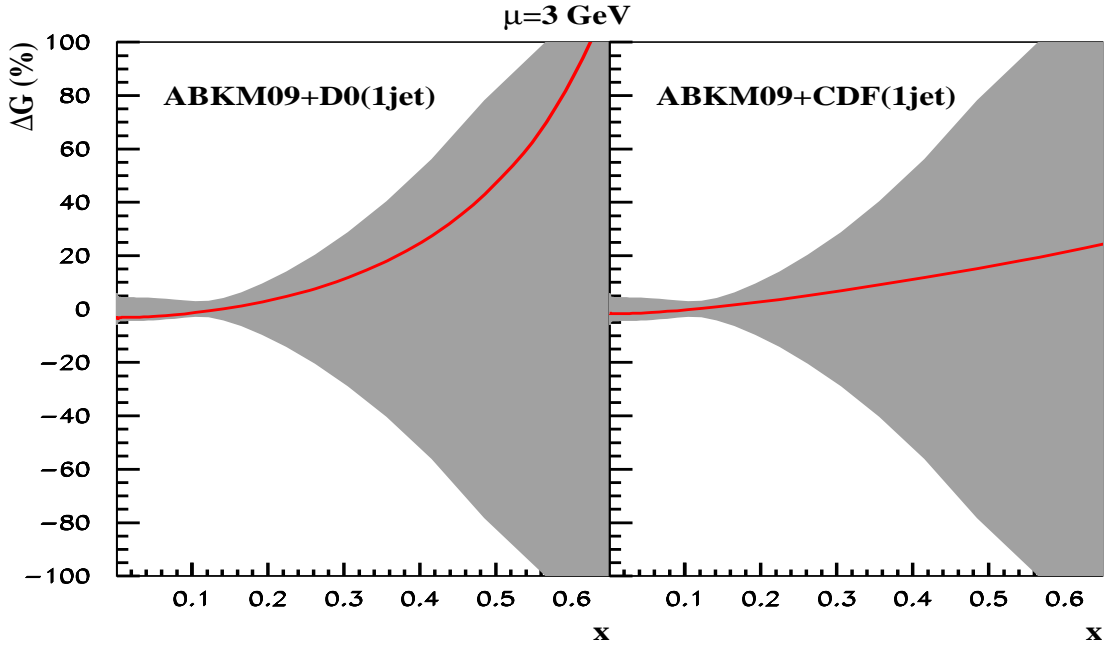}
\hfill
\caption{\label{fig:normglu} Relative variation of the gluon distribution 
due to dropping the normalization error in different Tevatron jet 
data sets (lines) compared to the 
uncertainties in the ABKM09 gluon distributions (shaded area) at 
the factorization scale $\mu=3~{\rm GeV}$ versus $x$
(left panel: D0; right panel: CDF). 
}
\end{figure}
To make an explicit check of the impact of the D0 
normalization uncertainty on the extracted PDFs we perform one more variant 
of the ABKM09 fit, with the normalization uncertainty in the D0 data dropped.
It turns out that dropping this error does not lead to any essential
deterioration of the D0 data description. 
For the variant of fit without the D0 normalization 
uncertainty taken into account  
the value of $\chi^2$ grows by less than 1 for 110 data points.  
The change in the gluon distribution obtained from these two 
variants of the fit generally does not exceed its uncertainty, 
cf. Fig.~\ref{fig:normglu}, and for other PDFs it is even smaller. 
This shows that the normalization error does not play  
crucial role in the interpretation of the D0 inclusive jet data. 
This can be also understood qualitatively, since
the normalization error in the data is 6.1\% only, much smaller than 
other systematic uncertainties, 
\begin{figure}[tbp]
\centering 
\includegraphics[width=.9\textwidth,height=0.8\textwidth]{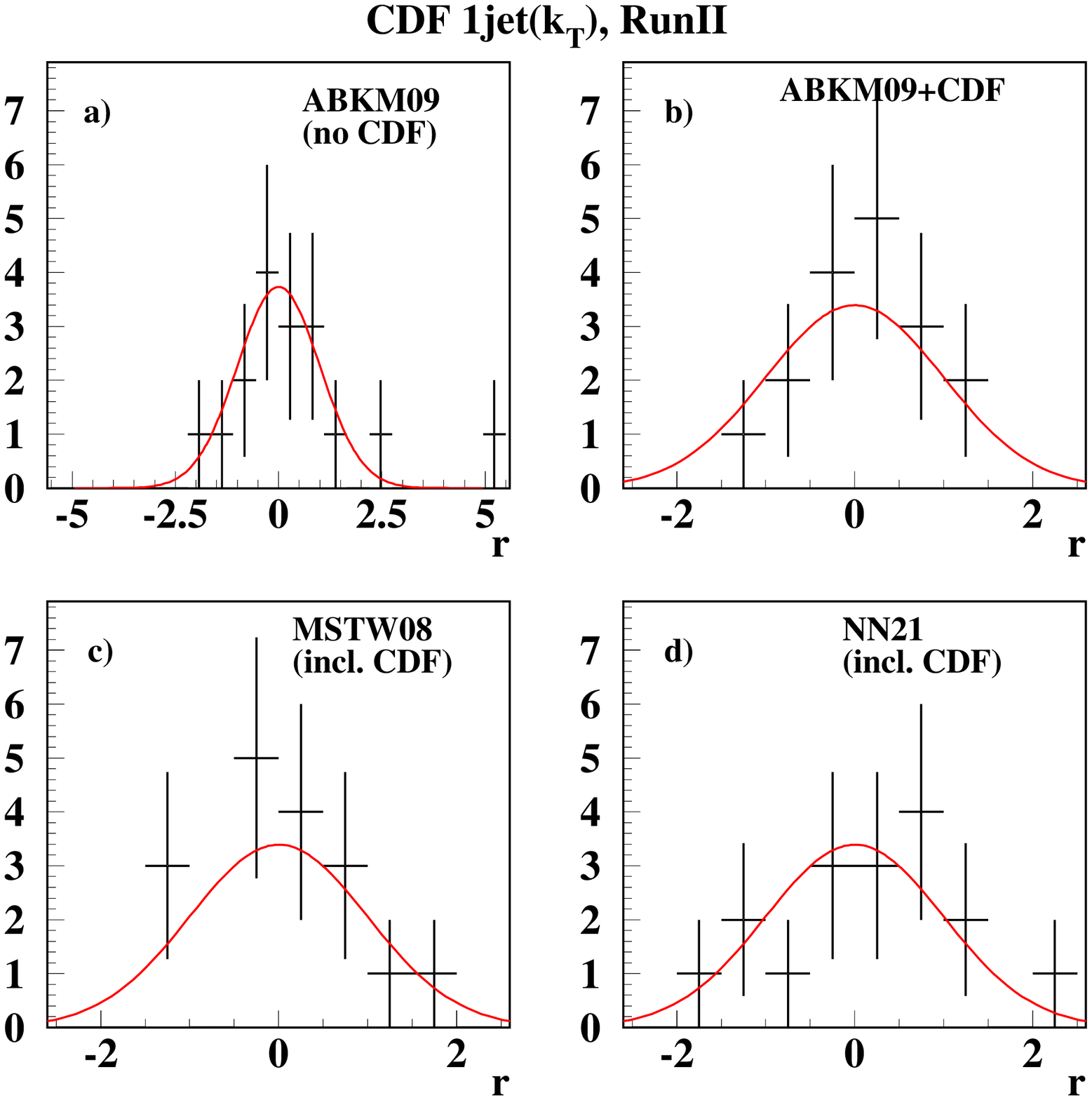}
\hfill
\caption{\label{fig:nuiscdf} The same as Fig.~\ref{fig:nuisd0} for the 
CDF data on inclusive jet production~\cite{Abulencia:2007ez}.
}
\end{figure}
therefore the latter easily overwhelm the impact of the normalization error. 

The CDF data on the inclusive jet cross sections~\cite{Abulencia:2007ez} 
were obtained with the $k_T$ algorithm for the jet definition and cover 
the range of $-2.1\div 2.1$ in the jet rapidity and $50\div 600$ GeV
in the transverse momentum of jet. The correlated systematic uncertainties in
the CDF jet data stem from 17 sources including the overall normalization. 
\begin{figure}[tbp]
\centering 
\includegraphics[width=.9\textwidth]{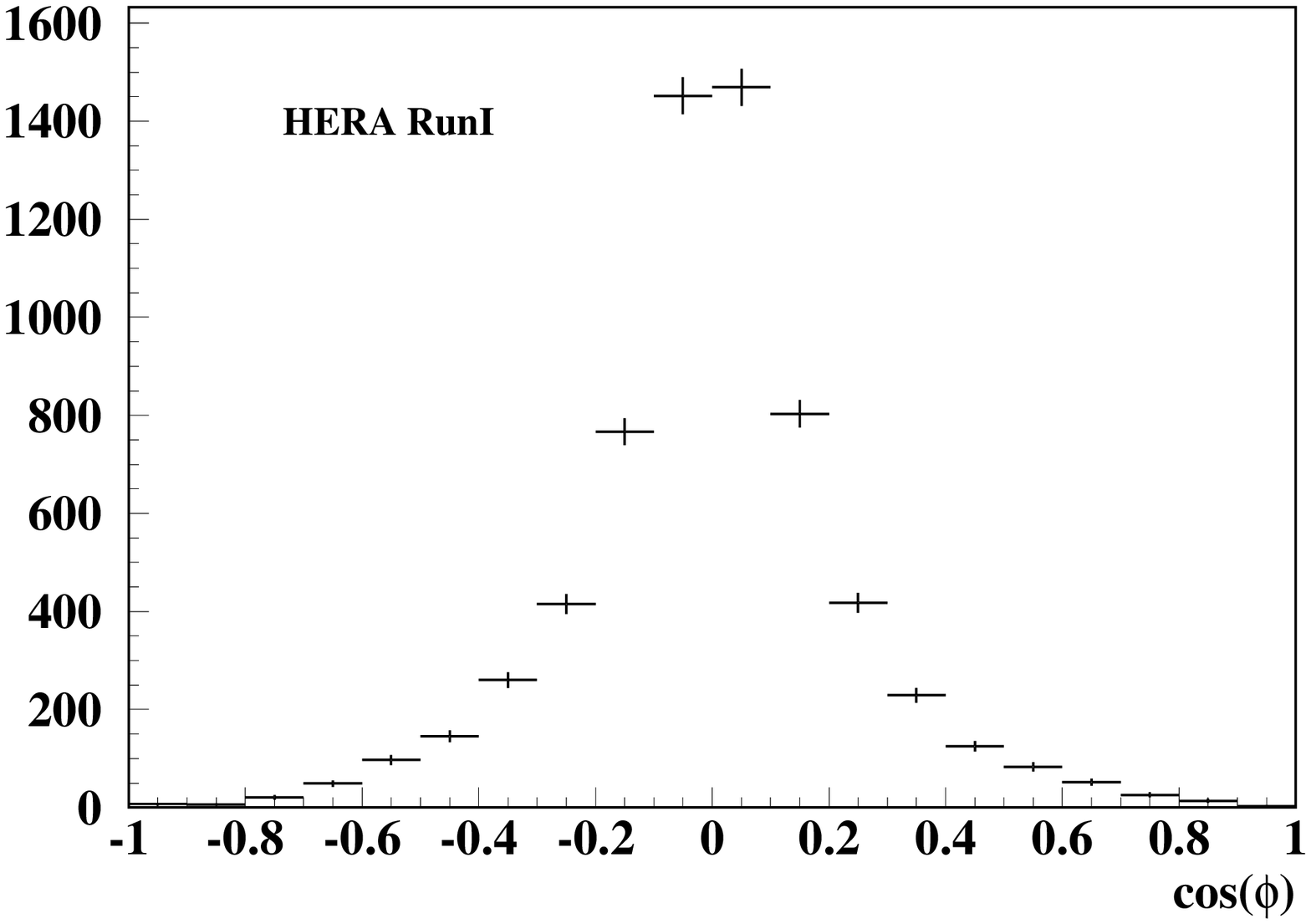}
\hfill
\caption{\label{fig:phiher} Distribution of the cosine of the angle between 
the systematic error vectors $\phi_{k k^\prime}$, cf. Eq.~(\ref{eq:phi}),
for the HERA data on the inclusive DIS structure 
functions~\cite{Aaron:2009aa}.
Only the angles with $k>k^\prime$ are histogrammed.
}
\end{figure}
\begin{figure}[hbp]
\centering 
\includegraphics[width=.9\textwidth,height=.5\textwidth]{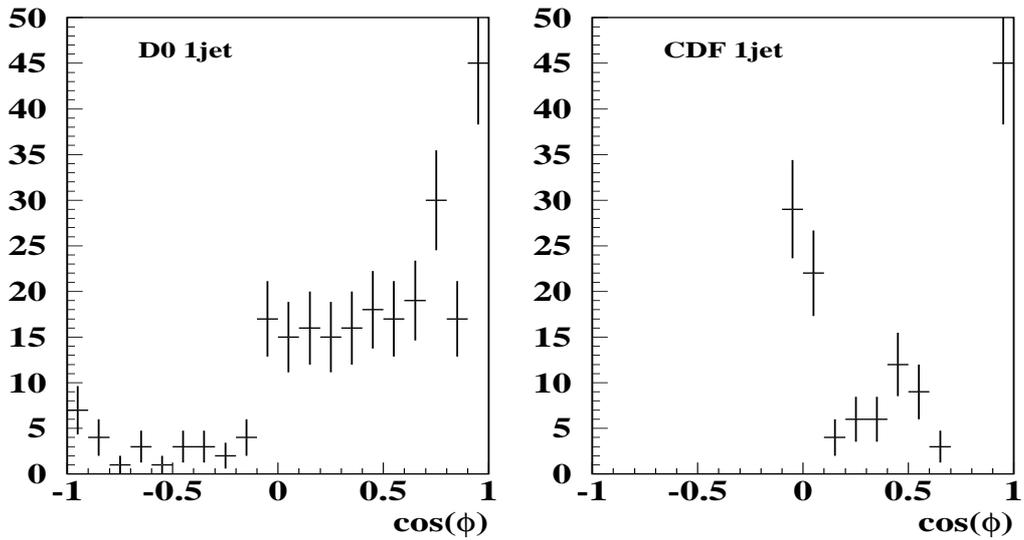}
\hfill
\caption{\label{fig:phitev} The same as in Fig.~\ref{fig:phiher}
for the D0~\cite{Abazov:2008ae} (left) and CDF~\cite{Abulencia:2007ez} (right) 
data on the inclusive jet production cross section. 
}
\end{figure}
The distribution of the corresponding nuisance parameters  
calculated with the NNLO ABKM09 PDFs is displayed in Fig.~\ref{fig:nuiscdf}. 
In general, it is in agreement with the normal Gaussian one with the 
only essential excess observed for the normalization nuisance parameter, 
which reaches the value of $r_{norm}=5.4$. This is bigger than the 
D0 normalization nuisance parameter. However, due to bigger uncertainties 
in the CDF data the errors in this parameter are also bigger as compared to the 
D0 case. The variance of the CDF normalization nuisance parameter
is $V(r_{norm})=0.93$
(to be compared to $0.85$ for D0) and the uncertainty due to the PDFs is 
$\Delta^{PDF}( r_{norm})=1.43$ (to be compared to 0.95 for D0). 
The CDF error due PDFs is evidently enhanced due to the particular
trend of the data with respect to the predictions based on the ABKM09 fit.
In the D0 case the offset of data does not depend on the jet energy, 
while the CDF jet energy dependence is systematically tilted as compared 
to the predictions, cf. Figs.~1,2 in Ref.~\cite{Alekhin:2011cf}.
With account of these errors the CDF normalization nuisance parameter is 
consistent with 0 within 3 standard deviations. 
Other nuisance parameters 
for the CDF data are consistent with zero within uncertainties,
cf. Fig.~\ref{fig:errnd0}, therefore 
in total the statistical significance of the excess in the normalization 
nuisance parameter is marginal, as well as for the D0 jet data. 
In line with this observation the distribution 
of the CDF nuisance parameters 
in the variant of the ABKM09 fit, which includes the CDF data, is in  
agreement with the normal Gaussian one, cf. Fig.~\ref{fig:nuiscdf}.
Similarly to the D0 case the change in $\chi^2$ due to dropping 
the CDF normalization uncertainty in the variant of the ABKM09 fit, which 
includes the CDF data, is marginal, i.e. less than 1 for 76 data points.  
The change in the PDFs due to dropping the normalization uncertainty is
also marginal, cf. Fig.~\ref{fig:normglu}. 

These observations do not support the conclusion of Ref.~\cite{Thorne:2011kq} 
about the crucial significance of the normalization uncertainty 
in the accommodation of Tevatron jet into the ABKM09 fit. As an explanation 
of this disagreement we point out that in the analysis of 
Ref.~\cite{Thorne:2011kq} 
the errors in nuisance parameters due to the PDF uncertainties and the   
experimental errors in the data are not considered. This leads to an  
overestimation of the statistical significance in the nuisance 
parameter excesses in the analysis of Ref.~\cite{Thorne:2011kq}.  
Another concern about the conclusion of Ref.~\cite{Thorne:2011kq} 
is related to the relevance of a rigorous statistical 
treatment of the systematic uncertainties in the Tevatron jet data.
Commonly, the
different sources of systematics are assumed to be independent, 
cf. Eqs.~(\ref{eq:data},\ref{eq:dataadd}). This also was assumed in 
the present study and in Ref.~\cite{Thorne:2011kq}.
We have checked this hypothesis for the Tevatron jet data 
plotting the cosine of angles between the systematic uncertainty
vectors $s_{k,i}$, which are defined as 
\begin{equation}
\cos(\phi_{k k^\prime})=\frac{\sum_i  s_{k,i}  s_{k^\prime ,i}}
{\sqrt{\sum_i  s_{k,i}^2 \sum_i s_{k^\prime ,i}^2}}.
\label{eq:phi}
\end{equation}
Naively, the distribution of $\cos(\phi_{k k^\prime})$ should
peak at $\cos(\phi)=0$ and be symmetric with respect to this peak 
for the case of independent sources of the systematic uncertainties.
In particular, such a picture is observed for the HERA data on the 
inclusive deep-inelastic-scattering (DIS) structure functions, 
cf. Fig.~\ref{fig:phiher}. However,
this is not the case for the D0 and CDF data, cf. Fig.~\ref{fig:phitev}. 
For both CDF and D0 data the distributions peak at $\cos(\phi)=1$ and are 
quite asymmetric, particularly in the case of CDF. This indicates 
a strong collinearity of many systematic uncertainty vectors. 
In case these systematic errors really stem from 
one of a few sources only,
the PDF fits based on the Tevatron jet data should be revisited.
Note that the vectors $s_{k,i}$ corresponding to the 
normalization uncertainty 
are collinear to many other systematic error vectors for these data. 
Evidently, this also explains the big error in the normalization parameter since
the corresponding nuisance parameters are mixed due to this collinearity. 

\begin{figure}[tbp]
\centering 
\includegraphics[width=.9\textwidth,height=.5\textwidth]{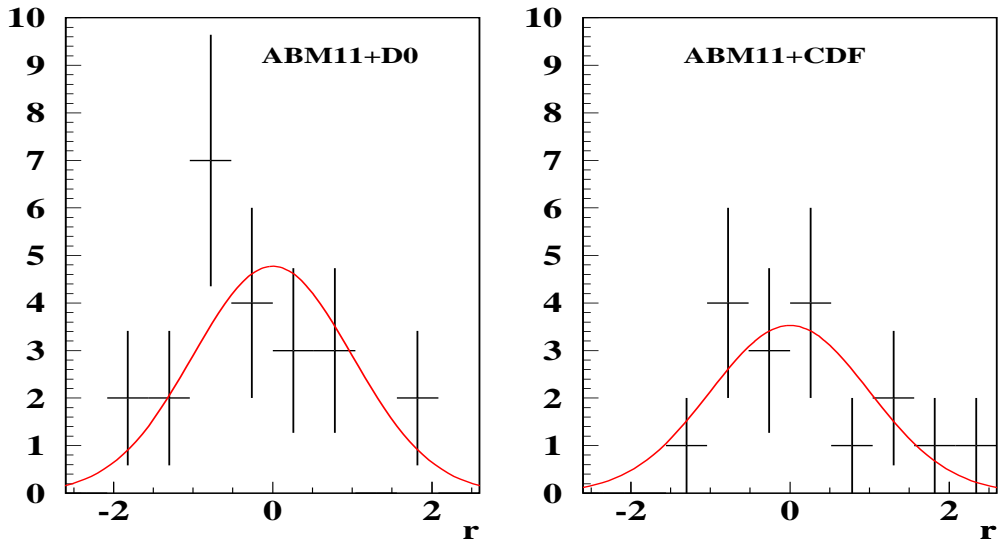}
\hfill
\caption{\label{fig:nuisabm11} The same as Fig.~\ref{fig:nuisd0} for the 
D0 data~\cite{Abazov:2008ae} (left) and 
the CDF data~\cite{Abulencia:2007ez} (right) and the variants of 
NNLO ABM11 fit~\cite{Alekhin:2012ig} including the D0 and CDF jet data,
respectively.}
\end{figure}

The distributions of the D0 and CDF nuisance parameters for the
variants of NNLO ABM11 fit~\cite{Alekhin:2012ig}, which include the Tevatron 
jet data in a similar way to Ref.~\cite{Alekhin:2011cf}, are in 
agreement with ones for the ABKM09 fit, cf. Fig.~\ref{fig:nuisabm11}.
In turn, both ABKM09 and ABM11 nuisance parameter distributions 
are similar to the ones obtained with the 
MSTW08~\cite{Thorne:2008am} and NN21~\cite{Ball:2011uy} PDFs, which are also 
tuned to the Tevatron jet data, cf. 
Figs.~\ref{fig:nuisd0} and \ref{fig:nuiscdf}. 
The remaining differences can be explained 
by the specific data selection in the fits and the fitted model 
peculiarities, like e.g. heavy-quark treatment, high-twist contributions,
and others, cf. Ref.~\cite{Alekhin:2012ig}. It can also appear due to  
different statistical estimators used in the PDF fit. In particular, 
the ABKM09 and ABM11 fits are based on the covariance matrix estimator of 
Eq.~(\ref{eq:chi2}), while in the MSTW08 fit the one of Eq.~(\ref{eq:chi2w}) is 
employed. As we have pointed out in Section~\ref{sec:basics}, in the first case 
the systematic errors are considered as multiplicative and in the second 
case as additive. 
Note, that an additive treatment of the errors in the cross sections 
leads to a statistical bias in the fitted parameters (cf. 
Refs.~\cite{D'Agostini:1993uj,Alekhin:2000es,JimenezDelgado:2012zx} 
and references therein for a discussion). Therefore 
it may have an impact on the nuisance parameter values which depend on the 
fitted PDF parameters as well. In the NNPDF 
fit~\cite{Ball:2011uy,Ball:2010de} the normalization errors are treated in a
special way, which allows to minimize the bias.
However, the covariance matrix of Eq.~(\ref{eq:cova}) is still used 
to take into account other correlated 
systematic errors (cf. Eq.~(1) in Ref.~\cite{Ball:2010de}). Since  
for the Tevatron jet data the latter dominate, the bias appears also in the
NNPDF fit.

\section{Conclusion}
\label{sec:concl}

We have analyzed a tension between the D0 and CDF  
inclusive jet data and the perturbative QCD calculations, which are based on the
NNLO ABKM09 and ABM11 PDFs with account of the NLO and NNLO 
threshold resummation corrections to the parton cross sections.
The nuisance parameters employed to quantify the tension
are calculated for each source of systematic uncertainty in the data
minimizing the $\chi^2$-estimator, which allows for shifts 
of the data by the value of systematic error scaled with the 
corresponding nuisance parameter. 
For some sources, in particular for the normalization uncertainty, 
the nuisance parameter values are relatively big. 
However, the analysis of their uncertainties due to the data fluctuations 
and the PDF errors shows that 
the nuisance parameter errors are as well substantial. 
In particular, this happens due to many systematic uncertainty vectors
including the normalization ones
being collinear and, as a result, the corresponding nuisance 
parameters are mixed. 
In view of those big uncertainties the statistical 
significance of the excesses in the normalization nuisance 
parameters is marginal. 
Furthermore, this conclusion is explicitly checked by considering
 the variants of  
ABKM09 fit, which include the Tevatron jet data without any normalization 
uncertainty taken into account. The results of these fits are quite 
similar to the ones including the normalization uncertainties.
These observations do not support the conclusion 
about the crucial role played by the normalization uncertainty in 
the accommodation of the Tevatron jet data into the ABKM09 fit 
mentioned in~\cite{Thorne:2011kq} disregarding the 
nuisance parameter errors. 
Besides, the statistical analysis of Ref.~\cite{Thorne:2011kq} lacks rigor
since the nuisance parameters 
are derived for the statistically biased estimator, while the ABKM09 fit 
is based on the estimator, which is asymptotically 
unbiased~\cite{Alekhin:2000es}.
At the same time, despite a serious statistical issue does not appear in the 
variants of the ABKM09 fit including the Tevatron jet 
data~\cite{Alekhin:2011cf}, 
the latter are finally not yet used in the ABM11 fit~\cite{Alekhin:2012ig} 
in view of yet lacking complete NNLO 
corrections, which may have an impact both on determination
of the strong coupling constant and on the parton distribution functions.

\acknowledgments

This work has been supported in part by DFG
Sonderforschungsbereich Transregio 9, Computergest\"utzte Theoretische
Teilchenphysik, and EU
Network {\sf LHCPHENOnet} PITN-GA-2010-264564.

\end{document}